Nonmetallic Gasket and Miniature Plastic Turnbuckle Diamond Anvil Cell for Pulsed Magnetic Field Studies at Cryogenic Temperatures

D.E. Graf[1], R.L. Stillwell[1,2], K.M. Purcell[1,2,3] and S.W. Tozer[1]

[1]The National High Magnetic Field Laboratory, Tallahassee, FL 32310, USA
[2]Florida State University, Department of Physics Tallahassee, FL 32306, USA
[3]University of Southern Indiana, Dept. of Geology and Physics, Evansville, IN 47712, USA

Abstract
A plastic turnbuckle diamond anvil cell (DAC) and nonmetallic gasket have been developed for pulsed magnetic field studies to address issues of eddy current heating and Lorentz forces in metal cells. The plastic cell evolved from our Ø 6.3 mm metal turnbuckle DAC that was designed in 1993 to rotate in the 9 mm sample space of Quantum Design's MPMS. Attempts to use this metal DAC in pulsed magnetic fields caused the sample temperature to rise to T>70 K, necessitating the construction of a nonconductive cell and gasket. Pressures of 3 GPa have been produced in the plastic cell with 0.8 mm culets in an optical study conducted at T = 4 K. Variations of the cell are now being used for fermiology studies of metallic systems in pulsed magnetic fields that have required the development of a rotator and a special He-3 cryostat which are also discussed.

Keywords: nonmetallic gasket, miniature plastic turnbuckle diamond anvil cell, fermiology, pulsed magnetic fields, extreme condition research

Introduction
The extreme conditions that must be created to study quantum critical phenomena such as competition between superconductivity and magnetism often require the use of pulsed magnetic fields for high pressure angular dependent studies. Typically these experiments are carried out at He-3 temperatures (350 mK<T<1.4 K) in a restrictive geometry resulting in a sample space not much greater than 10 mm. A typical 65 T capacitively driven pulsed magnet has a rise time of approximately 5-7 ms. This rapid change in magnetic field strength with time ($dB/dt$) leads to eddy current heating in any metallic components of a diamond anvil cell which is rapidly conducted through the diamonds to the sample, making the temperature of the sample indeterminate. This same issue is often overlooked in high pressure studies at millikelvin temperatures in dc magnetic field studies, where the relatively small amount of heat generated from a much smaller $dB/dt$ can cause the sample to heat from a base temperature of 20 mK to 200 mK or higher. Vibrations, induced by the coupling of the metal cell to the rapidly changing field in a pulsed magnet, can also result in a poor signal to noise ratio. If the metal components could be replaced with nonconductive material, the large thermal conductivity of the diamonds could be used instead to heat sink the sample to the cryogenic bath, allowing any eddy current heating of the sample to be more effectively dissipated. We have developed just such a nonmetallic cell based on our miniature metal turnbuckle design [1]. This 1.5 gram metal turnbuckle cell (fig. 1), constructed using BeCu and measuring Ø 6.3 mm x 9 mm in length, has been used extensively for dc magnetic field electrical transport studies [2-5] and *un*successfully in 1995 to study the organic quasi 2D superconductor (TMTSF)$_2$PF$_6$ in pulsed fields of 50 T. There was clear evidence of sample heating in the high pressure 4-probe electrical transport study which used a technique similar to that discussed in Ref. 6, but where particular attention was paid to the lead layout to minimize open loop pickup, another experimental challenge in pulsed magnetic field studies. A subsequent photoluminescence study that utilized a modulated GaAs/AlGaAs quantum well as a thermometer clearly showed that the sample experienced a temperature excursion from 4.1 K to temperatures in excess of 70 K during a full field 50 T pulse. Modeling indicated that even a stainless steel gasket reduced down to a washer 1 mm in diameter with a 0.5 mm hole and only 50 µm in thickness would drive the sample temperature to 8 K which necessitated the development of the nonconductive DAC. Several design concepts for the DAC were considered and tested, but the high symmetry of our original turnbuckle cell lent itself to rotation in magnetic fields and in the available sample space. It also proved to be the best design to minimize stress in the plastic components. The high symmetry of the turnbuckle design has also been useful in ongoing high pressure ESR [7], MAS NMR [8], single crystal x-ray (fig. 1) and high field SQUID studies.

Turnbuckle Design Concept
Turnbuckles are ubiquitous, being found in buildings and suspended walkways, bracing cattle gates or anchoring telephone poles where in every instance they are used in tension. Twisting the body of the turnbuckle draws together opposing cables, truncated with left and right hand threaded eyelets. Our DAC uses the turnbuckle concept to drive two anvils, captured on the inside of identical eyelets (endcaps, fig. 2), together to generate the pressure. Sapphire balls and anvils and moissanite anvils have also been used, but the focus of this paper will be on the diamond anvils as they are well suited to pulsed field studies. The turnbuckle diamond anvil cell must address the issues of translational and tilt alignment of the stones and also provide a means to drive the stones together while keeping them rotationally fixed so that the sample or anvils are not damaged. In our design, careful machining of the cell and carefully polished stones (a high degree of concentricity between the culet and girdle and parallelism between the culet and table) make the former possible. The latter is accomplished with spanner pins that fix the endcaps to the external load mechanism (fig. 3). The DAC can be made quite small, since this load mechanism is removable, the left and right-handed endcaps only being used to take up the load that is applied by this external mechanism.

A typical DAC might have anvil support tables for translational and tilt alignment and associated screws [9], a two part body, alignment pins and screws to apply and lock in the load. The turnbuckle DAC has only three parts plus the two anvils: the body (buckle) and the two endcaps (eyelets) (fig. 2). Much like the Mao-Bell cell [10] that relies on a long piston engagement length to diameter to ensure proper performance, cylindrical surfaces on the leading and trailing edges of the turnbuckle endcaps engage the body of the turnbuckle at its mid and end points, providing a reasonable length to diameter ratio that ensures proper alignment of the stones. This permits a loose thread engagement that allows for a free vertical displacement of 25-50 µm that is critical to the proper operation of this DAC. The external load mechanism can then advance the endcaps into the body a sufficient distance without engaging the threads of the body so that the body can then be turned with minimal torque, locking in the new load. Clocked blunt threading allows for synchronized advancement of endcaps. For our miniature metal turnbuckle, an English 10-48 thread (48 threads per inch) was used. This gives adequate thread strength while still producing a reasonably fine resolution for anvil displacement, since the anvils advance together at twice the thread pitch, a disadvantage of this design [11].

No epoxy or retainer ring is used to set the stones in the endcaps, a light press fit with the walls of counter bored recesses provides a mechanically robust setting that thermally cycles without failure. The stones are set into these recesses by tacking the culet to one end of a high precision vise that is stood on end; lightly closing the vise to assure that the stone is square. The retractable jaw with stone affixed is then withdrawn until the endcap can be slid onto the fixed jaw and located under the stone. The bezels of the stone and a lead taper on the counter bored recess enable the stone to be pressed into and be seated firmly on the bottom of the recess. In addition, the counter bore is scalloped, leaving only three narrow points of contact with the stone girdle to minimize the load needed to set or remove the stone. Optical access is through a 0.65 mm diameter hole in each of the endcaps. The cell can hang from the end of a 600-micron fiber optic when the Teflon jacket of the fiber optic is screwed into the 00-90 threaded counter bore at the termination of this optical access port. It is by this means that the cell is loaded into the MPMS SQUID magnetometer, the fiber being brought through a miniature vacuum quick connect. The fiber thus serves the dual purpose of drive shaft and means to calibrate the pressure at the experimental temperature as the pressure in any DAC changes as a function of temperature. Three equally spaced, milled holes around the perimeter of the endcap's backside (fig. 2) accept the spanner pins from the two platens of the loading apparatus. These platens are themselves pinned together with larger diameter lapped dowel pins that assure parallelism and torsional rigidity. This fixes the endcaps with respect to one another so that the stones can move only in and out and not spin on the gasket, which would result in sheared sample leads or contacts. A load from three large bolts (which we have recently incorporated into the large lapped pins) or an external source such as a hydraulic press pull or push these two large plates together and advance the endcaps into the turnbuckle body. A split metal collar clamps around the cell body and is used to provide leverage to lock in the load (Originally, the cell body was machined with a hex so that a wrench could be used to turn the body, but we found that this led to over torquing and cell failure.). A fiber optic is brought through the load apparatus and to the anvil surface of the DAC so that the pressure can be monitored *in situ* while the load is being changed. In many ways, the external load mechanism resembles a Merrill-Basset cell [12] with the turnbuckle DAC replacing the diamond seats. Three additional holes, spaced around the perimeter of each endcap, provide a means to bring wires into the cell for 4-probe transport, Hall effect or skin depth measurements that are discussed later in this paper. Any of these holes that are not utilized for leads provide a path for cryogenic fluid when the sample space is tight. Slots in the platens of the load cell, aligned with the through holes in the endcaps, provide a means to bring the electrical leads out of the endcap, preventing them from being damaged. CNC machining capabilities have enabled the manufacturing of 50 cells in the period of two weeks. It has also allowed us to experiment with pin geometries and we have found that a pin defined by an elongated arc is much more robust than a pin with circular cross-section.

Plastic version of the turnbuckle DAC
A variety of plastics and other nonconductive materials ceramics such as alumina and yttria stabilized zirconium, commercial composites such as G10 and phenolics and natural composites such as warthog tusk were tried as a construction material for the body and endcaps of the plastic cell. The final choice, Parmax 1200 [13], is a bulk unreinforced polymer with a ultimate tensile strength of 203 MPa, a compressive strength of 351 MPa and good machining characteristics [14]. Under applied load it cycles well to cryogenic temperatures if cooled slowly. Care must be taken not to use solvents to clean the DAC when under load as these tend to creep into microcracks in the plastic and cause catastrophic failure. Recent attempts to enhance the properties of Parmax by introducing nano particles have shown great promise [15], increasing the bulk modulus in thin films by a factor of two. We are exploring this modified Parmax as a potential gasket material or material for the cell, if it can be made in bulk form. Although a standard 60° included angle vee thread is used for our metal turnbuckles, we have adopted the BA thread with radiused root and crest for our plastic cells to minimize stress risers.  The thread in the endcaps and the thread relief in the cell body are the weak points in the design. This is due to the machined finish that can be achieved in this plastic which leaves small pits in the surface that act as crack nucleation sites in the tapered thread where it terminates.  Lapping of the threads with a diamond paste improves the surface finish.  We are also experimenting with annealing the cell parts to eliminate this problem.  The current solution, a secondary machining operation removes the lead thread half way around the circumference leaving a blunt thread.  In addition to providing a means to advance the endcaps evenly, the blunt thread on the lead edge of the endcap and cell body eliminates a stress riser that often caused cell failure in earlier versions.  The maximum load that the larger (denoted EC17, a later, more user friendly version of EC15) cell can support is 4 kN.  For the plastic cells, the split collar that turns the cell body is sleeved with a thin split G10 tube.  This cushions the plastic and the glass fibers in the G10 provide more friction to grip the body than the all-metal collar used for the metal cells.  The optical access hole in recent versions has been changed to a 20° included angle cone for better optical observation and for orienting and monitoring mosaicity of the crystal *in situ* via x-rays.

Nonmetallic Gasket

The non-metallic gaskets are based on the composite metal/diamond epoxy gasket that Boehler [16] developed wherein he removes the indented region of a metal gasket and replaces it with a heavily loaded diamond epoxy mix.  We use Emerson & Cummings 1266 or Loctite Hysol E120HP epoxy and a 50:50::1 µm:0.5 µm diamond powder mix.  The powder is blended with a drop of epoxy using a mortar and pestle.  The mix is ready when it has a flaky consistency and is highly reflective.  The diamond epoxy mix is pressed to the correct thickness between two mandrels with tapered ends and flats that mimic the included angle and culet of the diamonds to be used [fig 4].  For our completely non-metallic gasket, we have replaced the portion of the metal gasket that surrounds the indented region with Zylon HM fiber [17] that has been reduced to a quarter tow width and wetted with Loctite Hysol E120HP epoxy.  The epoxy impregnated Zylon is secured to a yoke that rigidly aligns and fixes the mandrels and is wound under tension around the diamond epoxy disk taking the shape defined by the tapered tips of the mandrel.  The finished assembly is heated at 65 °C for 2 hours to cure the epoxy. Breaking the finished gasket free of the mandrels requires that the mandrels be highly polished (mandrel are made from A2 drill rod hardened in a vacuum environment) and coated with a release agent.  Loctite Frekote sealer and release is a good solution, but we have also used sprayed Teflon and bee's wax with limited success in the past.  The sample hole is then drilled or milled with tungsten carbide tooling. We have also used mandrels with holes drilled into the tips that are of a size appropriate for the sample hole and connected the two holes with a gage pin yielding a gasket that is ready to use. The advantage of this latter setup is that the holes are dead center, but we have had issues with obtaining a clean release since the epoxy has a tendency to flow into gaps between the gage pin and hole in the mandrel. Other groups have improved on the density of the diamond epoxy mix by pressing it prior to cutting out the disk and by using a small amount of solvent to thin the epoxy [18]. The included angle on the mandrels is 5° larger than that of the stones used and the mandrel tip is 10-20 μm larger than the culet so that the diamond pavilions do not load the Zylon. This makes it possible to use smaller loads to obtain the desired pressure; however, exaggerating these dimensions much more than this causes the diamond epoxy material to extrude out and tear the leads. A very thin ring of 5-minute epoxy is run around the perimeter of the right-hand endcap's diamond culet and run down the pavilion to secure the gasket to the stone and provide an initial leak tight seal.

Pressure medium
Experiments to this point have been carried out using Fomblin 140/13, Daphne 7373 [19], Fluorinert 70/77, a 4:1 mix of methanol and ethanol [20] or glycerine as the pressure medium. While pressure losses of 0.1-0.3 GPa upon cooling to 4 K are typical for the plastic turnbuckle DACs we find that the actual amount depends on the value of the starting pressure, the pressure medium used and the volume taken up by the sample and its compressibility. Interestingly, a metal version of EC17 increases in pressure almost 20% upon cooling to cryogenic temperatures.

A gear driven load cell was developed for cryogenically loading argon. When cryoloading, one starts with the sample and ruby trapped in the empty gasket and only opens the cell, via the gear drive, to allow liquid in after the cryogens have settled, thus minimizing the chances that the sample will slip out of the hole when the diamonds are apart and the violent action of the cooling liquid can sweep the sample out. After repeated attempts, however, it was found that the diamond epoxy used in the gasket was too brittle to make this a practical means of loading Ar. A device has recently been constructed that makes use of a high pressure intensifier [21] to load gases at room temperature, but it has yet to be tested.

Procedure for loading and changing pressures in the turnbuckle DAC when using room temperature liquids for pressure media

To load the cell, push the right-hand endcap onto the three spanner pins of one of the platens that make up the load cell. Each endcap and the ends of the turnbuckle body should be marked with letters that make it clear which is right and left-handed. This becomes critical when the cell is assembled and it is not possible to see the thread. By convention, we have established that the gasket will be fixed to the right-hand endcap. All sample preparation work can now be done with the endcap secured to this platen. An easily seen mark should be made on the side of this platen that indicates the start of the endcap's blunt thread. The platen should also be marked with "RH" which will help during the loading and unloading. Once all preparation work is complete and the pressure medium and ruby have been added, the turnbuckle body is screwed onto this endcap with the split collar tightened onto the body. One should check that the spanner pins are fully engaged into the endcap by advancing the pins using the 4-40 set screws that support them from the underside. Place the left-hand endcap onto the other platen of the load cell and mark the start of the blunt thread as done for the RH endcap. Check for full pin engagement and align the blunt threads of the RH and LH endcaps. Insert the large lapped pins into the RH platen of the load cell and slip the LH platen onto these pins, gently bringing the LH endcap to the body of the cell. With the load cell on its side, lightly press the platens together by hand and back the turnbuckle body off the RH endcap. Once a low click is heard, indicating that the body is unthreaded from the RH endcap, listen for another click from the LH endcap. This should not require a rotation of more than 5° if the blunt threads have been aligned. While still maintaining light pressure on the load cell assembly, turn the split collar in the opposite direction which will now pull the endcaps together. Sufficient pressure on the load cell must be maintained during this operation to prevent one of the endcaps from pulling away from the spanner pins and stripping any wires that might be protruding from the cell. If it does not appear that the cell is pulling together symmetrically, start over, as most likely one of the endcaps did not engage on initial take-up. When one feels the cell seal, hand tighten the body while pressing on the spanner assembly before putting it into the press. Place the RH platen against the bottom of the press, the plates of which have been previously aligned. This convention allows one to remember which way to turn the body in order to tighten or loosen the DAC. Increase the load in small increments to take out the 25-50µm play in the thread displacement while continuously tightening the body until one sees the load on the gage drop slightly. A piezo load cell works nicely for this purpose. It is advisable to log the loads and the torques used to lock in the load so that, when it comes time to release the pressure, one can reproduce the required load and locking torque. We find that it is most useful to run a fiber optic down to the cell so that the pressure can be known at all times using the fluorescence from a ruby chip [22].

To unload or change the pressure in the cell, place the split collar lined with split G10 sleeve onto the turnbuckle body of the loaded cell and tighten the 4-40 screws that clamp it to the cell body. Trial and error will determine what is tight enough. A small strip of tape affixed to turnbuckle body and the split spanner and then slit makes a good telltale in verifying that the collar is tight enough and not spinning on the body. Locate the letters "R" and "L" scribed on the bevel of the DAC's endcaps or on the ring of the turnbuckle body just outside the bevel of the cap. Push the right-hand endcap onto the load cell platen labeled "R". Push the spanner pins from the backside to make certain that they are all fully engaged. Insert the large lapped pins that are shared by the RH and LH platens. Gently push the spanner pins in the LH platen into the left hand endcap and verify their engagement. As per convention, place the RH platen on the bottom plate of a press. Initially calibrating the pressure with the ruby manometer will allow the load from the press to be increased until it is seen that the cell pressure increases, at which point it becomes possible to turn the cell body and raise or lower the pressure. It may be necessary to put a rod into one of the holes drilled into the perimeter of the split collar to provide sufficient torque.

Preparation for transport studies
A thorough discussion of the tuned tank circuit technique based on the tunnel diode oscillator [TDO] for pulsed fields can be found in reference 23 and a more general discussion is given by Van der Grift [24, 25]. Condensed to its simplest form, the TDO is a driven LC tank circuit where any changes in the sample's resistance, caused by pressure, temperature or magnetic fields, is detected by a coil that surrounds the sample and is read as a change in the frequency of the circuit. With care, a noise level of tens of Hertz can be realized making it possible to see Shubnikov de Haas oscillations.

Electrical leads (twisted pairs or flexible coax such as Gore's CXN 3369 which is used for the TDO measurements) are brought in through one or more of the three through holes equally spaced around the perimeter of the endcap. These leads are anchored to the endcaps with epoxy. A small length of thin Kapton tubing acts as a boot to stress relief the leads where they exit the endcap. To minimize the open loop area and noise due to vibrations, the leads are brought through closely spaced holes in the Zylon portion of the gasket and secured to the gasket. Coated copper wire (7-25 μm in diameter) or silver plated BeCu wire are used to make the small 3-10 turn coils used in the contactless TDO studies and these are brought out over the diamond epoxy disk in parallel grooves that were created using short lengths of tungsten wires that were pressed into the disk. The cross sectional area of the impression is kept smaller than that of the Cu or BeCu wires so that they completely fill the impression when the diamonds press them down under load, thus guaranteeing a good seal while at the same time minimizing the occurrence of sheared leads.

There are several advantages to using the TDO technique in pulsed magnetic fields when compared to the conventional 4-probe technique. For time sensitive samples such as radiological specimens it is quite easy to place a sample into the coil. This also minimizes exposure for the student loading the cell. Some samples also form oxide layers that make it difficult to establish ohmic contacts while others react with the metal used in many of the conductive paints. Making contacts that are ohmic, small, of minimal resistance and robust to cryogenic cycling and pressure excursions is also an art unto itself. A TDO setup eliminates these issues. With careful consideration given to how the coil leads are brought into and out of the gasket hole it is also possible to minimize open loop pickup. With some care a figure-of-eight coil can also be wound which almost completely eliminates this contribution to the signal. In addition, the TDO arrangement does not stress the sample unlike that which it experiences when tethered to leads.

Cryogenics and Rotation
A pulley driven rotator has been made from the same material as the cell. It is based on the design of Palm and Murphy [26] and uses the Russian spring described by Eremets [9], but rendered in PEEK plastic (this allows the spring to be located at cryogenic temperatures close to the rotator) to keep the pulley string drive under tension. The spring is concentric to the He-3 tail. To minimize eddy current heating and vibrations, the He-3 tail and vacuum can are made with 12 μm thick fiberglass cloth impregnated with epoxy that is spirally wound onto a mandrel coated with a release agent similar to that used for the gaskets, cured, and pulled off to yield a very thin wall cryostat tail. Caps for these tails are fabricated from G10 and the 100 μm gap is maintained by small pieces of 100 μm Nylon fishing line that is tacked to the He-3 tail using Emerson and Cumming 1266 epoxy. Using thin wall stainless steel tails for the He-4 cryostat with minimal gaps between the tails, it is possible to create enough sample space to rotate the EC15 DAC at He-3 temperatures in a pulsed magnet with a 77 K bore of 15.5 mm.

Fermi surface study
Using the methods described in this paper, an initial high pressure study of $CeIn_3$ [27] found a phase transition at high magnetic fields identified as a

Lifshitz transition [28] that collapsed with pressure onto the Néel phase boundary. Subsequent studies using clean carefully annealed samples have let us study the evolving Fermi surface of this material, the 0.1 GPa results for which are shown in fig. 5.

Acknowledgements
This work was funded by DOE NNSA DE-FG52-10NA29659, NSF DMR-0654118, the State of Florida and DOE DE-PS52-05NA26772. Technical expertise was provided by S. Outland, J. Farrell, R. Desilets, R. Newsome and R. Schwartz. Eddy current calculations were performed by Y. Eyssa. High purity CeIn$_3$ crystals used in this study were grown and annealed by K.N. Collar and J. Bourg (technique taught by J.C. Cooley, Los Alamos National Laboraory) as part of an undergraduate research project.

References

1. The idea to use a turnbuckle in compression for use as a diamond anvil cell came about from a hike in Colorado with Dieter Hochheimer in the fall of 1993, where we came across a large turnbuckle tensioned power transmission pole which caught the eye of SWT as it was out-of-place in the meadow. A BeCu turnbuckle DAC was produced within a few weeks of returning to the lab and was tested in the QD's MPMS where we were able to resolve Hc$_2$ in a superconductor, but did not have the sensitivity to see dHvA oscillations. Magnetometry work requires a magnetically clean alloy be used for the cell. For a long time it was thought that the cobalt in the standard BeCu C17200 alloy was responsible for the increase in susceptibility at low temperatures, but J. Cooley and M. Aronson, "Origins of paramagnetism in beryllium-copper alloys", Journal of Alloys and Compounds **228** 195-200 (1995) found that the background signal is instead related to the iron impurities in the material. For subsequent turnbuckle DACs, we had Brush Wellman produce a magnetically clean alloy of BeCu:Co that has a low temperature susceptibility close to that of binary BeCu, but possesses mechanical properties equivalent to the standard C17200 BeCu:Co alloy. As with all high pressure cells, the pressure limit is governed by the anvil material, culet area, size and alignment of the anvils, sample volume and compressibility, cell and gasket geometry and choice of pressure medium. To date a maximum pressure of 8 GPa has been generated in a BeCu version of this cell with 10-48 threads without anvil failure using 1 mm 16-sided culet, 0.22 carat diamonds and a solution of methanol:ethanol:water::16:3:1 as a pressure medium. A strikingly similar design was recently published by G. Giriat,

W.W. Wang, J.P. Attfield, A.D. Huxley and K.V. Kamenev, "Turnbuckle diamond anvil cell for high-pressure measurements in a superconducting quantum interference device magnetometer", Review of Scientific Instruments **81** 73905-1 to 73905-5 (2010).
2. P. Goddard, S.W. Tozer, J. Singleton, A. Ardavan, A. Bangura and M. Kurmoo, "Angle-dependence of the magnetotransport and Anderson localization in a pressure-induced organic superconductor", Synthetic Metals **137** 1287-1288 (2003).
3. A.F. Bangura, P.A. Goddard, S. Tozer A.I. Coldea, R.D. McDonald, J. Singleton, A. Ardavan and J. Schleuter, "Angle-dependent magneto-transport measurements on κ-(BEDT-TTF)$_2$Cu(NCS)$_2$ under pressure", Synthetic Metals **153** 449-452 (2005).
4. N. Kurita, M. Kano, M. Hedo, Y. Uwatoko, J.L. Sarrao, J.D. Thompson and S.W. Tozer, "Investigation of YbInCu4 at pressures to 7 GPa", Physica B **378-380** 104-106 (2006).
5. M. Kano, H. Mori, Y. Uwatoko and S. Tozer, "Anisotropy of the upper critical field in ultrahigh-pressure-induced superconductor (TMTTF)$_2$PF$_6$", Physica B **404** 3246-3248 (2009).
6. S.W. Tozer and H.E. King, Jr., Electrical Transport Measurements on Fragile Single-Crystals to 7.5 GPa in the Diamond Anvil Cell, Review of Scientific Instruments **56** 260-263 (1985).
7. C.C. Beedle, S.W. Tozer, C. Morien and S. Hill, to be published.
8. L. O'Dell, I.L. Moudrakovski, D.D. Klug, C.I. Ratcliffe, S.W. Tozer, M. Kano, and S. Desgreniers, in preparation.
9. M.I. Eremets, High-Pressure Experimental Methods (Oxford University Press, Oxford, 1996).
10. H.K. Mao and P.M. Bell, "Design and Varieties of the Megabar Cell", Carnegie Inst. Washington Yearbook **77**, 904-908 (1978).
11. We have experimented with 72 pitch threads for endcaps made from MP35N or NiCrAl, but find that a BeCu body must be used to avoid galling. Current versions of the metal cell are machined from metal in the heat treated (aged) state to negate distortions and magnetic oxides that form on the surface of the two Ni alloys.
12. L. Merrill and W.A. Bassett, "Miniature Diamond Anvil Cell for Single-Crystal X-ray Diffraction Studies", Review of Scientific Instruments **45** 290-294 (1974).
13. Parmax has had an interesting history as a high performance material. Originally made by Dow Chemical as a potential competitor to Dupont's Vespel, it was spun off to Maxdem who then built Mississippi Polymer Technology to try and bring it up to quarter pilot quantities. Hurricane Katrina leveled MPT, the formulation rights being sold to Solvay who sold molding rights to Ensinger Plastics. Production recently ceased due to issues with material quality, but it is hoped that Parmax (now marketed as Tecamax SRP120) will be reintroduced in 2013.

14. V.J. Toplosky, R.P. Walsh, S.W. Tozer and F. Motamedi, "Mechanical and Thermal Properties of unreinforced and reinforced polyphenylenes at cryogenic temperatures", Advances in Cryogenic Engineering (Materials) **46**, Eds. U.B. Balachandran, D.U. Gubser, K.T. Hartwig and V.A. Bardos, Kluwer Academic/Plenum Publishers (2000).
15. C-Y. Chang, E.M. Phillips, R. Liang, S.W. Tozer, B. Wang and C. Zhang, "Alignment Interaction and Properties of Carbon Nanotube Buckypaper / Liquid Crystalline Polymer Composites", to be submitted to European Polymer Journal.
16. R. Boehler, private communications and R. Boehler, M. Ross and D.B. Boercker, "Melting of LiF and NaCl to 1 Mbar: Systematics of Ionic Solids at Extreme Conditions", Physical Review Letters 78 4589-4592 (1997).
17. Toyobo now produces Zylon although it was originally made by Dow Chemical around the time that Dow formulated Parmax. Zylon is now used extensively in the reinforcing shells for pulsed magnets.
18. N. Funamori and T. Sato, "A cubic boron nitride gasket for diamond-anvil experiments", Review of Scientific Instruments 79 053903-1 to 053903-5 (2008).
19. K. Yokogawa, K. Murata, H. Yoshino, and S. Aoyama, "Solidification of high-pressure medium Daphne 7373", Japanese Journal of Applied Physics, Part I, **46** 3636-3639 (2007).
20. G.J. Piermarini, S. Block and J.D. Barnett, Journal of Applied Physics **44** 5377-5382 (1973).
21. R.L. Mills, D.H. Liebenberg, J.C. Bronson and L.C. Schmidt, "Procedure for Loading Diamond Cells with High-Pressure Gas", Review of Scientific Instruments **51** 891-895 (1980).
22. J.D. Barnett, S. Block and G.J. Piermarini, Review of Scientific Instruments **44** 1-9 (1973).
23. T. Coffey, Z. Bayindir, J.F. DeCarolis, M. Bennett, G. Esper and C.C. Agosta, "Measuring radio frequency properties of materials in pulsed magnetic fields with a tunnel diode oscillator", Review of Scientific Instruments **71** 4600-4606 (2000).
24. C.T. VanDegrift, "Tunnel-Diode Oscillator for 0.001 PPM Measurments at Low-Temperatures", Review of Scientific Instrument **46** 599-607 (1975).
25. C.T. VanDegrift and D.P. Love, "Modeling of Tunnel-Diode Oscillators", Review of Scientific Instruments **52** 712-723 (1981).

26. E.C. Palm and T.P. Murphy, "Very low friction rotator for use at low temperatures and high magnetic fields", Review of Scientific Instruments **70** 237-239 (1999).
27. K.M. Purcell, D. Graf, M. Kano, J. Bourg, E.C. Palm, T. Murphy, R. McDonald, C.H. Mielke, M.M. Altarawneh, C. Petrovic, R. Hu, T. Ebihara, J. Cooley, P. Schlottmann, and S.W. Tozer, "Pressure evolution of a field-induced Fermi surface reconstruction and of the Néel critical field in $CeIn_3$", Physical Review B **79** 214428-1 to 214428-7 (2009).
28. P Schlottmann, "Lifshitz transition with interactions in high magnetic fields", Physical Review B **83** 115133 (2011).

Figure Captions

Fig. 1. From left to right, the original Ø6.3 mm metal turnbuckle DAC with an endcap placed on the coin for reference, a variation for x-ray beamline studies which has a 4-Omega included angle of 82° on the upstream and downstream sides, two variations of the plastic DACs (EC15 and EC17 with a sectioned turnbuckle and endcap above the latter) and the EC15 DAC in the plastic rotator used for pulsed field studies at He-3 temperatures. 3.0 mm girdle stones with an overall height of 1.65 mm are used in the smaller EC15 plastic DAC, which has an outer diameter of 9.0 mm. Culet diameters vary between 0.6 and 1.0 mm with the overall height being kept constant by varying the girdle height. This cell rotates within a He-3 tail having a diameter of 11.4 mm. 4 mm diameter girdles provide more support and subsequently higher pressures in the larger Ø11.4 mm EC17 DACs. The EC15 and rotator are made from original Parmax stock which was a clear yellow. As the Parmax was scaled up for market, impurities and new processes, yielded a darker brown and finally an opaque purple stock.

Fig. 2. Sectional view of the plastic EC17 DAC and three views of an endcap showing the endcaps: *a* and *a+*; turnbuckle body: *b*; anvil: *c*; gasket: *d*; throughhole for leads: *e* and spanner hole: *f*. The sectional cut is through one of the lead holes and the opposing spanner pin hole. The two lapped surfaces on the turnbuckle body and endcaps that provide an effectively large engagement length to diameter are emphasized with red lines on the lower left of the DAC in the assembled view.

Fig. 3. The first panel shows various pieces of the load cell assembly with the turnbuckle endcaps placed on the spanner pins of the platens and the cell body held by the split collar. The large lapped dowel pins, which keep the platens parallel and tortionally rigid, are shown in the left platen for clarity. The second panel shows a sectional view of the assembled load cell with two of the six

spanner pins shown. For clarity, the collar is not shown. The last panel is a picture of the assembled load cell and turnbuckle DAC with half of the split collar removed to shown the slit G10 sleeve.

Fig. 4. The gasket jig (yoke, part *a*, and mandrels, part *b*) and, to the lower right, the resulting nonmetallic gasket which is shown on the far right fixed to an anvil with a figure-of-eight coil inside the gasket hole. The expanded view of the mandrels in the middle shows how the resulting shape mimics a conventional metal gasket. The hoop strength of the Zylon HM (high modulus) fiber wet wound around the diamond epoxy disk gives the support that would normally be provided by the extruded region of a conventional metal gasket. The diamond epoxy disk varies in thickness between 60-180 μm, depending on sample and coil size and the required pressure range.

Fig. 5. A downsweep field trace for $CeIn_3$ (c-axis parallel to field) taken at 1.4 K that shows the change in frequency of the TDO circuit. The total signal has contributions from the magnetoresistance of the coil wire and the sample. The dashed black line is a fit to the data. The minimum at 56 Y is attributed to the Lifshitz transition. The field induced Néel transition cannot be seen in this trace as the available fields were too low. *Insets*: The resulting background subtracted data set that clearly shows Shubnikov de Haas quantum oscillations and the FFT of this data set showing an orbit with a frequency of 3350 T. A complete P-B-T-θ phase diagram that allows us to map out the reconstructed Fermi surface will permit a better understanding of the field induced Lifshitz and Néel transitions.

Fig.1

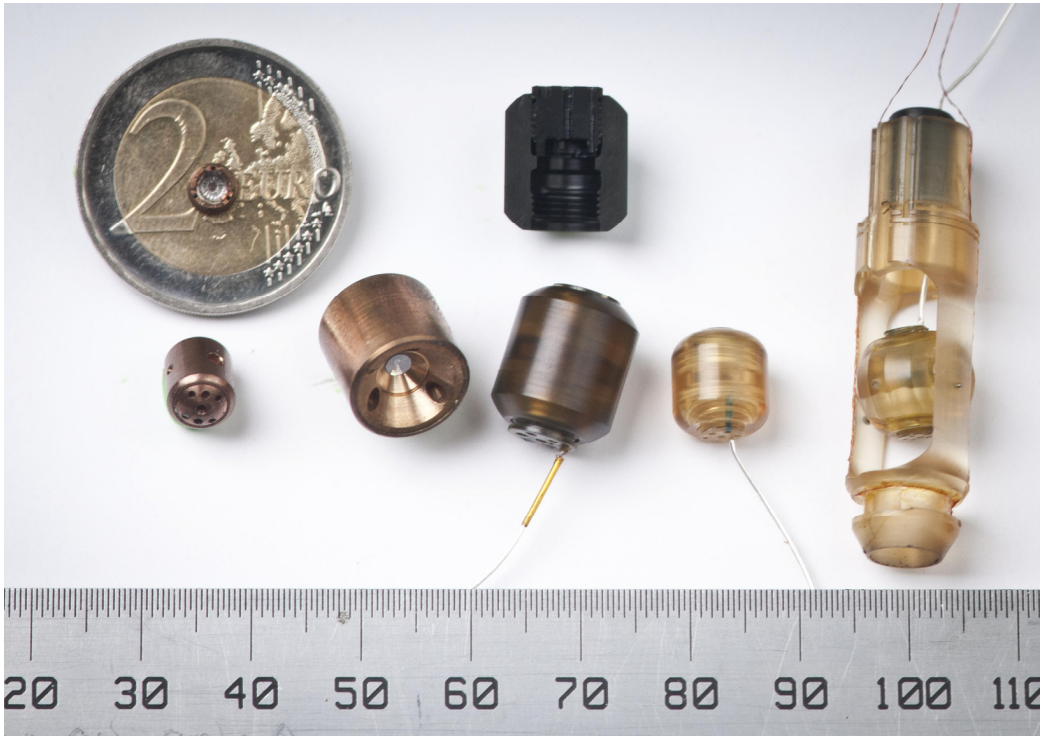

Fig.2

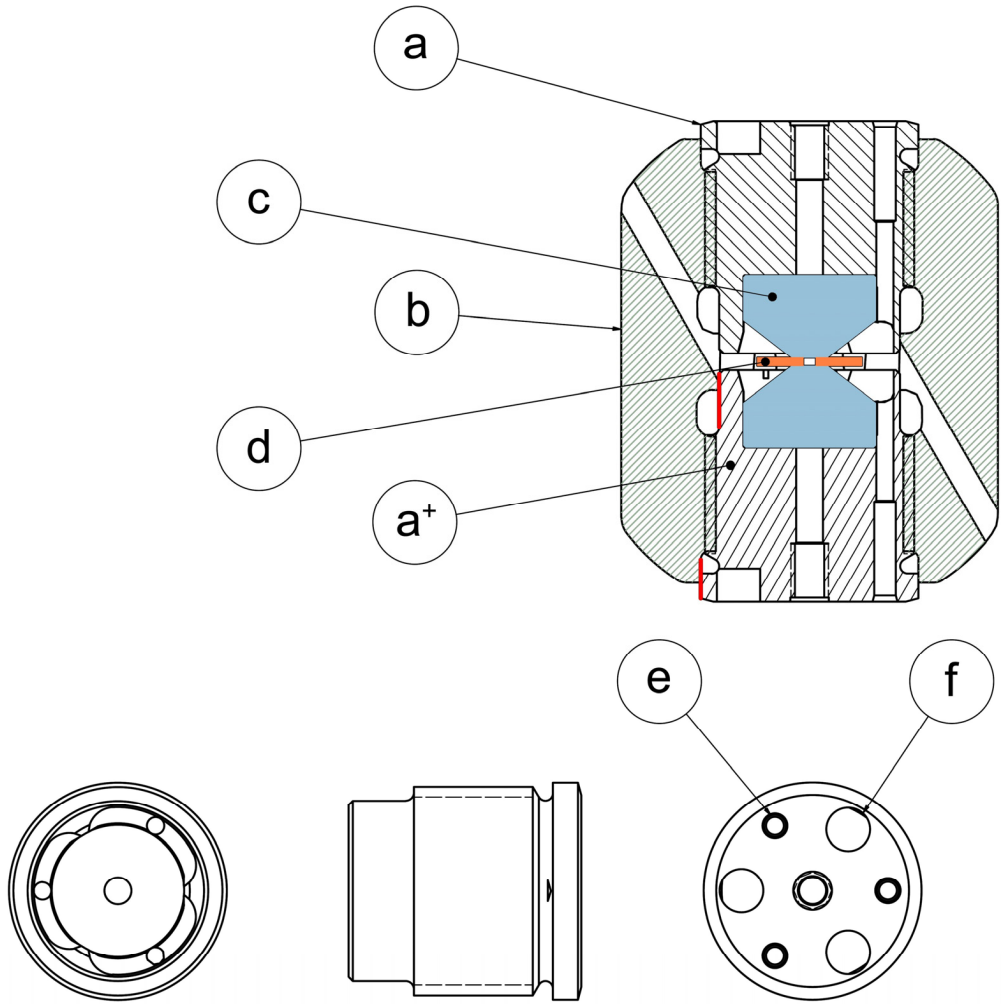

Fig.3

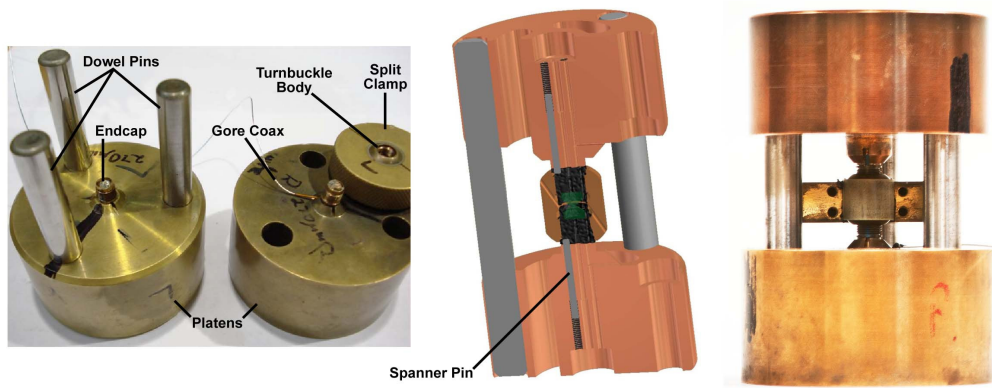

Fig.4

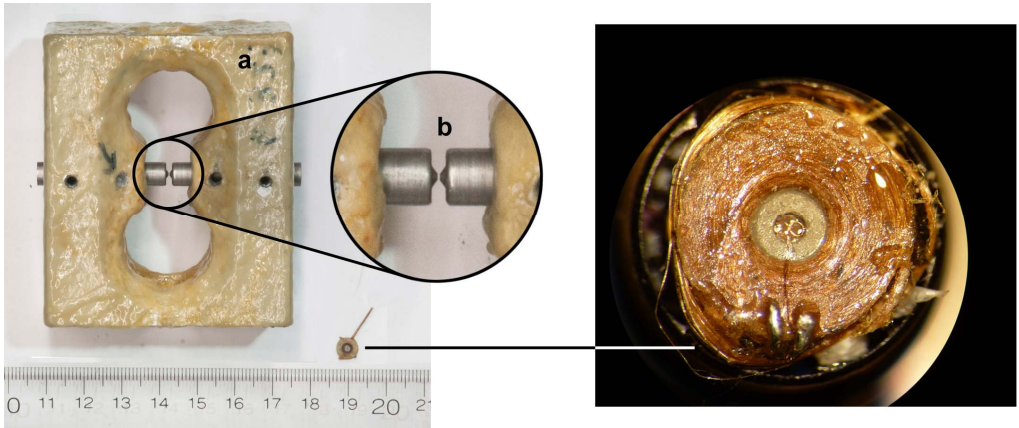
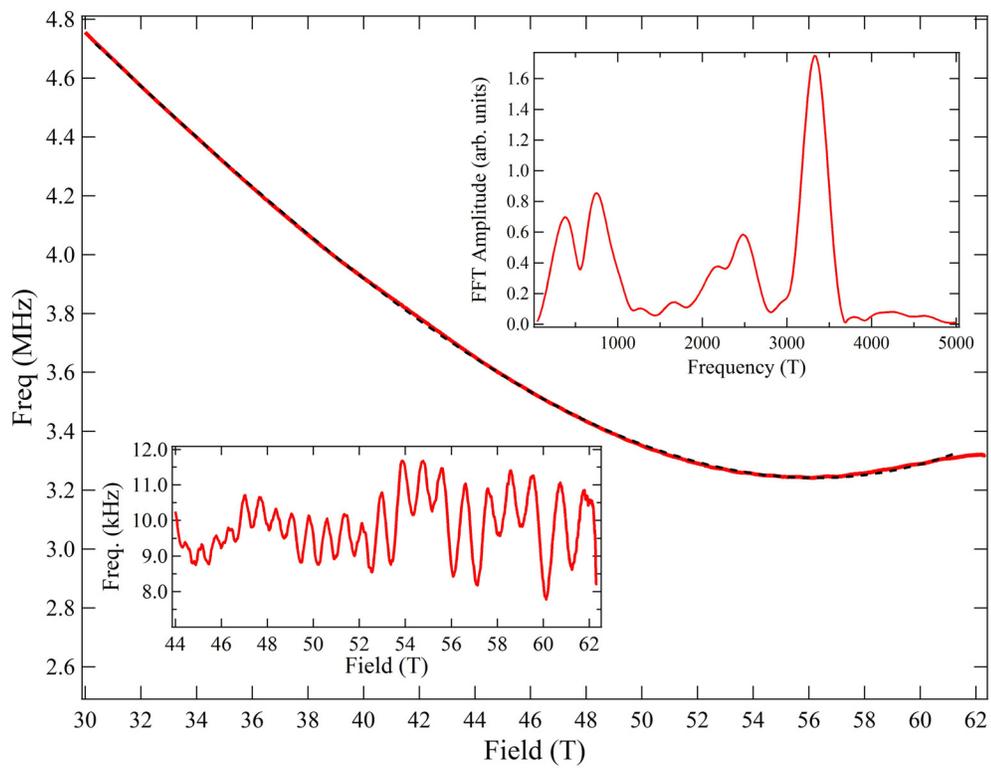

Fig.5